\documentclass{article}

\PassOptionsToPackage{numbers,compress,sort&compress}{natbib} 
\usepackage{graphicx}
\usepackage{amsmath} 
\usepackage{bbm}
\usepackage{newunicodechar}
\usepackage[dblblindworkshop, final]{neurips_2026}
\workshoptitle{ReMuCAI: Real-Time Multimodal Conversational AI}

\newunicodechar{−}{-}

\usepackage[utf8]{inputenc} 
\usepackage[T1]{fontenc}    
\usepackage{hyperref}       
\usepackage{url}            
\usepackage{booktabs}       
\usepackage{amsfonts}       
\usepackage{nicefrac}       
\usepackage{microtype}      
\usepackage{xcolor}         

\title{Multi-Party Backchannel Prediction: a Diagnosis, a Benchmark, and a Ceiling}

\author{%
  Mohammed HAFSATI \\
  Enchanted Tools \\
  \texttt{mohammed@enchanted.tools} \\
  \And
  Ahmed LOUGHZALI \\
  Enchanted Tools \\
  IMT Nord Europe \\
  \texttt{Ahmed.loughzali@etu-imt-nord-europe.fr} \\
}

\begin{document}

\maketitle

\begin{abstract}
Backchannel prediction has been studied almost entirely in dyadic conversation. We introduce a multi-party benchmark based on the AMI corpus \citep{carletta2007unleashing}, comprising 682 masked-listener views from 171 meetings, 190 speakers, and 18{,}697 backchannel events, with a person-disjoint held-out split. A state-of-the-art dyadic model applied zero-shot to meeting audio performs at chance (AUROC $0.499$), nevertheless its frozen acoustic features remain informative: a linear probe reaches $0.704$, and retraining the predictor raises performance to $0.751$. Retraining reveals a second limitation. Listener conditioning improves prediction for listeners seen during training but not for unseen listeners, and the gap remains under capacity reduction, listener-adversarial training, per-listener adaptation, and oracle lexical conditioning. Adversarial training removes only part of the speaker-identity information, while stronger removal hurts prediction, suggesting that identity is entangled with cues that are useful for backchanneling. A within-model control helps explain this pattern: with the same features and data splits, turn-onset prediction transfers to unseen listeners, while backchannel prediction does not. Backchannel rates also vary about twice as much across individuals as turn-onset rates. Since backchannels occupy only about $1\%$ of frames, frame-level F1 is strongly affected by the base rate. We therefore report AUROC alongside event-F1 on listener-active regions. We release the benchmark and evaluation tools\footnote{\url{https://github.com/HafsatiMohammed/bc_multiparty_release}}.
\end{abstract}

\section{Introduction} 
\label{sec:intro} 

Conversational fluency depends not only on what an agent says, but also on how it behaves while listening. Backchannels, brief vocalizations or gestures such as ``mm-hm,'' ``yeah,'' or a nod, signal attention and engagement. An embodied agent that remains silent throughout a long utterance can therefore feel unresponsive. Producing backchannels naturally requires predicting, in real time, when a response is appropriate. Voice Activity Projection (VAP) models \citep{ekstedt2022voice} make these predictions directly from raw audio and have been fine-tuned for backchannel prediction \citep{inoue2025yeah}. However, they have been studied almost entirely in dyadic settings with a clean, separate channel for each participant. Social robots often interact with groups, where the input contains a mixture of several speakers and any participant may take the role of listener. It remains unclear whether a model trained for dyadic interaction can transfer to this setting and, if not, what prevents it from doing so. We study this question using a benchmark built from the AMI meeting corpus \citep{carletta2007unleashing}. Each participant is treated in turn as a \emph{masked listener} standing in for the robot, and their backchannels must be predicted from the mixed audio of the remaining participants (Section~\ref{sec:benchmark}). Three main findings emerge. \textbf{The failure lies in the predictor, not the representation.} A strong dyadic model applied zero-shot to meetings performs at chance (AUROC\footnote{area under the receiver operating characteristic curve} $0.499$). Yet a linear probe trained on its frozen acoustic features reaches $0.704$, and retraining only the predictor raises performance to $0.751$ (Section~\ref{sec:localization}). \textbf{A cross-listener gap persists despite every remedy we test.} Prediction is consistently better for listeners seen during training than for unseen listeners. Listener conditioning, capacity reduction, listener-adversarial training, per-listener adaptation, and oracle lexical conditioning all fail to close this gap (Secs.~\ref{sec:conditioning} to \ref{sec:adaptation}). Speaker identity can only be partially removed from the representation, and stronger removal hurts prediction, suggesting that identity information is entangled with cues that are useful for backchanneling. \textbf{The gap appears to come from the behavior itself.} Using the same features, trunk, and data splits, turn-onset prediction shows no comparable gap. The difference between the two prediction heads is significant under a meeting-permutation null, while backchannel rates vary about twice as much across individuals as turn-onset rates (Section~\ref{sec:dissociation}). These results suggest that backchanneling is more person-specific than turn-taking and that the listener-agnostic acoustic component available for prediction is relatively small. Because backchannels occupy only about $1\%$ of frames, frame-level F1 is strongly affected by the base rate. We therefore rank systems by AUROC and report event-F1 on listener-active regions (Section~\ref{sec:eval}). We release the benchmark, data splits, and evaluation tools.

\section{Related work} \label{sec:related} 

We organize prior work along a speaker-listener axis and ask whether reported improvements come from personalization to listeners seen during training or from transfer to listeners not seen before.

\textbf{Speaker-side cues.} Most work treats backchannels as responses to local cues from the speaker, including prosody, pauses, utterance boundaries, gaze, and discourse structure \citep{ward2000prosodic,terrell-mutlu-2012-regression}. \citet{inoue2025yeah} formulate the task as continuous, causal, frame-level prediction by fine-tuning VAP on unbalanced data, and their model provides the methodological basis for our work. In our setting, the speaker's acoustics serve as the baseline, and we consider a listener effect meaningful only if it provides information beyond those cues. Visual signals also contain useful backchannel information \citep{morency2010probabilistic}, but reliable gaze is difficult to obtain in many practical settings, where participants may look at screens, notes, or one another and far-field capture can make tracking less reliable. We therefore focus on audio alone. 

\textbf{Listener idiosyncrasy.} \citet{Blomsma_Vaitonyte_Skantze_Swerts_2024} showed that fourteen addressees responding to the same controlled speaker stimulus differed in both backchannel frequency and form. Because the speaker stimulus was held constant, this provides strong evidence that backchannel behavior varies across listeners. However, it does not test whether a representation learned from some listeners can generalize to a person who was absent from training. 

\textbf{Identity and attribute conditioning.} \citet{ortega2023ohjeezuhhuhlisteneraware} add listener-ID embeddings to lexico-acoustic features on Switchboard and report modest improvements, while \citet{ortega2023modelingspeakerlistenerinteractionbackchannel} explicitly model speaker-listener interaction, increasing F1 from $.42$ to $.55$. On German GECO, however, listener-only conditioning performs worse than speaker-only conditioning. Both approaches rely on discrete ID lookups and conversation-level splits, so their gains mainly demonstrate personalization to known participants rather than transfer to unseen ones. \citet{park2024backchannel} replace IDs with conversational-pattern features on Korean counseling data, although a listener's own model is often not the best-performing model for that listener, and the person vector is derived from the conversation being predicted. \citet{fukunaga2025backchannel} instead use age, gender, and relationship type. These attributes transfer better than using no participant information, but perform worse than ID embeddings. This creates a clear trade-off: IDs capture more individual variation but generalize poorly, while broader attributes transfer more easily but capture less. This motivates both an audio-derived listener representation and, more importantly, a person-disjoint evaluation, which these studies do not report. 

\textbf{Multi-party backchanneling.} \citet{heylen-op-den-akker-2007-computing} studied 13 AMI meetings and predicted, given that a backchannel had occurred, which of three listeners produced it, achieving $51\%$ accuracy against a $36\%$ baseline. This is an event-conditional selection task rather than continuous prediction. \citet{10.1145/3527188.3561926} developed a continuous audio model for an artificial third-party listener observing a dyad. We therefore do not claim to introduce the first multi-party backchannel predictor. Our contribution is instead a controlled, person-disjoint evaluation of continuous prediction in natural meetings, together with an analysis of why transfer fails. Relatedly, \citet{cavalcanti2025dyadosyncrasy} find that turn-transition timing is more specific to a pair than to an individual. Since their work concerns turn-taking rather than backchanneling, we use it mainly to motivate a listener-speaker interaction hypothesis.

\section{Benchmark, task, and evaluation protocol} \label{sec:benchmark} 

\subsection{Masked-listener views and labels} 

AMI \citep{carletta2007unleashing} contains close-talk headset recordings of English meetings with three to five participants. For each meeting, we create one view per participant. That participant becomes the \emph{masked listener}, standing in for the robot; channel~1 contains their headset audio, while channel~2 is the mean of all remaining headsets and represents what the listener hears from the rest of the room. The resulting task keeps the same two-channel structure as dyadic backchannel prediction, but differs in two important ways that we examine throughout the paper. 

Channel~2 contains a mixture of several speakers rather than a single interlocutor, and the listener can be any meeting participant rather than a curated active backchanneler. This produces 682 views from 171 meetings and 190 speakers, totaling 399.5~h of audio. We divide them into 20-s windows of 400 frames at 20~Hz, the encoder frame rate described in Section~\ref{sec:model}, yielding 71{,}588 segments. 

Backchannel onsets come from AMI's gold \texttt{bck} dialogue-act annotations for the 139 meetings in which they are available. For the remaining 32 meetings, we use a rule that labels a listener utterance as a backchannel if it lasts at most 1.2~s, contains at most three tokens, matches a 16-entry lexicon, and occurs while another participant speaks within 0.5~s before and 1.5~s after it. 

On the gold meetings, this rule recovers annotated events with precision $0.634$ and recall $0.729$ (Appendix~\ref{app:benchmark}). After merging consecutive backchannels, the benchmark contains 18{,}697 events, of which 12{,}594 are gold and 6{,}103 are rule-based. 

Rule-labeled segments are used only for training and account for $40\%$ of the training set; every metric reported in the paper is computed on gold-annotated meetings. Turn onsets (\textsc{start}), defined as the onset of non-backchannel listener speech regions,\footnote{Listener speech regions are merged across gaps shorter than 0.25~s, so a \textsc{start} follows at least 0.25~s of listener silence.} are constructed in the same way and provide a within-model control in Section~\ref{sec:dissociation}. Training targets are shifted into the future as in VAP-BC (Section~\ref{sec:model}), while unshifted onset times are stored separately for event-level evaluation. 

\subsection{Splits} \label{sec:splits} 

Training, validation, and test meetings are disjoint, and validation listeners do not appear in any other split. We divide the test set into two portions. The \emph{unseen} portion contains six complete AMI participant groups, corresponding to 24 listeners across 24 meetings and 10{,}020 segments. None of these listeners appears in training or validation. The \emph{seen} portion contains 14 additional meetings with 56 participants who also occur in the training set, but in different meetings, for a total of 6{,}044 segments. The unseen portion is therefore strictly person-disjoint, while the seen/unseen comparison separates personalization to known listeners from transfer to new ones, a distinction that prior work does not isolate. The two test slices differ in backchannel prevalence by $1.5\times$ (Appendix~\ref{app:benchmark}). We therefore compare base-rate-dependent metrics across configurations only within the same slice and use AUROC for seen-versus-unseen comparisons. 

\subsection{Evaluation protocol} \label{sec:eval} 

Backchannels occupy only $1.37\%$ of valid frames in the gold test set, with 67{,}044 positives among 4{,}882{,}204 frames, or roughly 75 negatives for every positive. Precision and frame-level F1 are therefore strongly shaped by the base rate. Restricting evaluation to frames in which a backchannel is plausible, meaning that the listener is silent and someone else is speaking, helps much less in meetings than in dyadic data. Someone other than the listener is already speaking $73.6\%$ of the time, so this restriction increases the positive rate only to $1.65\%$. Requiring multiple concurrent speakers removes most of the data for only a small additional increase in density (Appendix~\ref{app:selectivity}). No natural slice of the multi-party data approaches the ${\sim}10\%$ backchannel density found in curated dyadic datasets, which explains much of the difference in absolute F1 relative to dyadic reports (Appendix~\ref{app:f1}). We therefore rank systems using area under the ROC curve (AUROC), which does not depend on the class base rate. This follows the same motivation as earlier VAP work, which moved from F1 to balanced accuracy under a milder imbalance \citep{ekstedt2022voice}. For comparability, we also report frame-level F1 over both all valid frames and listener-active frames, together with event-level F1 obtained by peak picking with a $\pm300$~ms collar. In both cases, the threshold is selected on validation data and then applied unchanged to test (Appendix~\ref{app:event_f1}). Because all listeners in the same meeting share the underlying recording, the listener split is also \emph{meeting}-disjoint. Seen and unseen test listeners come from 14 and 24 non-overlapping meetings, respectively, so any observed difference between them can reflect both listener novelty and meeting-to-meeting variation, which is substantial. To account for this, we evaluate every seen/unseen contrast against a \emph{meeting-permutation null}. We generate 20{,}000 size-matched reassignments of the seen/unseen label across meetings and recompute the statistic for each reassignment. Variation across random seeds measures reproducibility on a fixed collection of meetings and therefore cannot replace this test. We report seed-averaged estimates for consistency across experiments, but all significance claims are based on the meeting-permutation null.

\section{Model and training} 
\label{sec:model} 

\begin{figure}[t] \centering \includegraphics[width=0.65\linewidth,trim=300 120 320 120,clip]{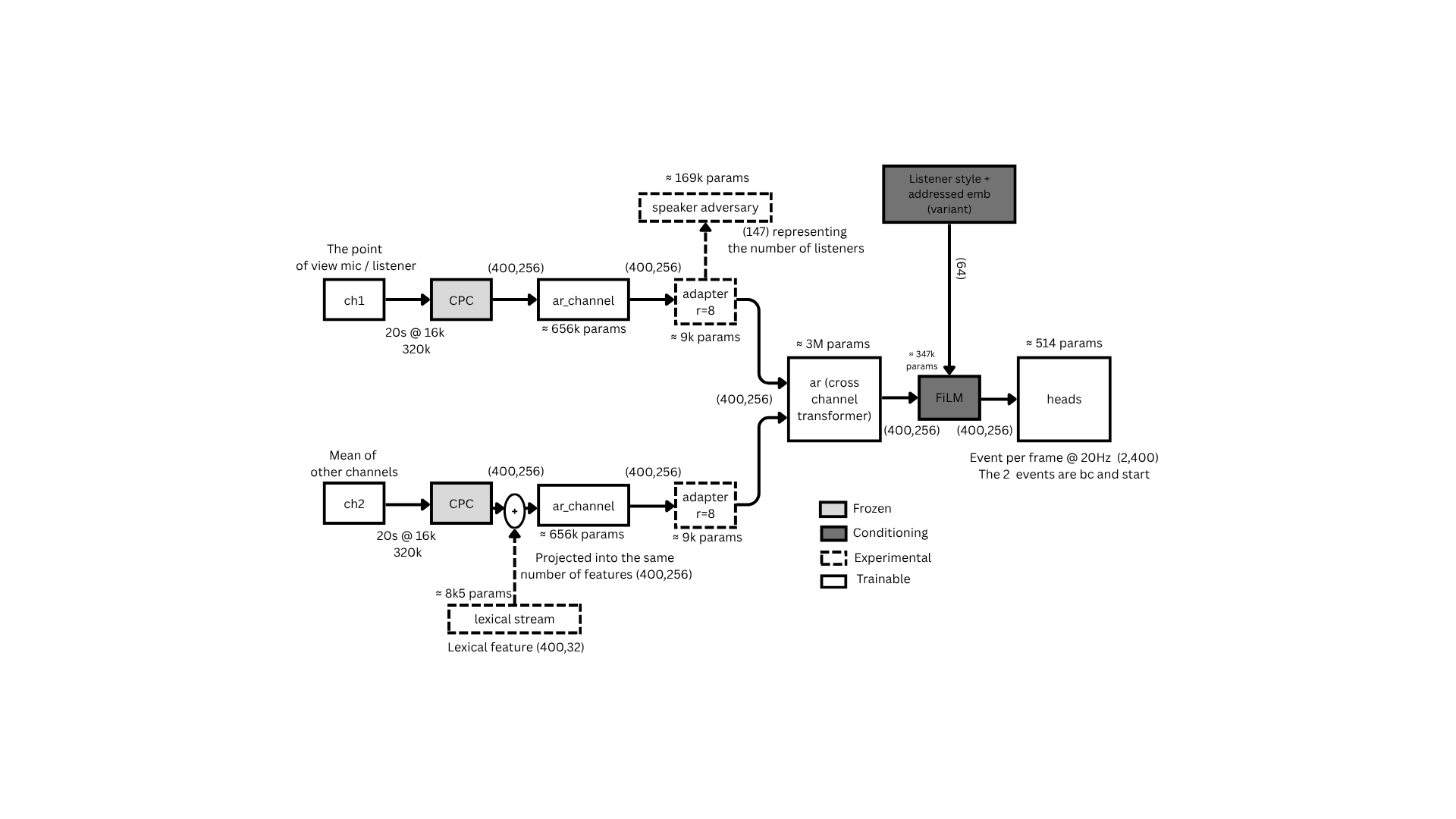} \caption{VAP-BC with the optional components evaluated in Section~\ref{sec:remedies}: listener/floor-holder FiLM conditioning, rank-8 adapters, listener-identity adversary, and lexical stream.} \label{fig:model} \end{figure}

We follow the VAP-BC architecture of \citet{inoue2025yeah} using the MaAI implementation\footnote{\url{https://github.com/MaAI-Kyoto/MaAI}}. Figure~\ref{fig:model} shows the model together with the optional components evaluated in Section~\ref{sec:remedies}. Both 16-kHz audio channels are passed through a frozen pretrained CPC encoder \citep{oord2018cpc,riviere2020cpc}, with weights shared across channels. This produces 256-d features at 20~Hz, giving $\mathbf{X}\in\mathbb{R}^{2\times400\times256}$ for each 20-s segment. 

Freezing the encoder is important for our analysis: any information used by the rest of the model must already be present in these features, which is precisely what the linear probe in Section~\ref{sec:localization} tests. A one-layer channel-wise transformer with 656K parameters, shared across channels, maps the two streams to $o_1,o_2\in\mathbb{R}^{400\times256}$. A three-layer cross-attention transformer with 2.89M parameters then combines them into $h\in\mathbb{R}^{400\times256}$. The full model has 8.0M parameters, or 8.3M with conditioning, of which 3.6M and 4.0M are trainable, respectively. 

\paragraph{Conditioning.} An optional amortized listener style vector $z\in\mathbb{R}^{64}$ modulates $h$ through FiLM \citep{perez2018film} immediately before the prediction heads: $h'_t=(1+\gamma(z))\odot h_t+\beta(z)$. We initialize $\gamma$ and $\beta$ to zero, so the conditioned model begins exactly as the unconditioned model and only learns to use listener information when doing so reduces the loss. In the \emph{floor-holder} variants, a frame-level embedding of the participant currently holding the floor is concatenated to $z$. We compare three conditioning settings, \emph{none}, \emph{listener}, and \emph{listener+floor-holder}, with channel~2 either containing the mixture of all other participants or only the current floor-holder's audio.

\paragraph{Targets and loss.} Linear heads applied to $h'$ (or $h$ in the unconditioned model) produce frame-level logits for backchannel onset (\textsc{bc}) and turn onset (\textsc{start}). Following VAP-BC, targets are shifted into the future: a frame is positive if an onset occurs within a horizon $H=1$~s (20 frames) after it. The checkpoint's original VAP/VA heads receive no gradient and are not used. Positives are rare ($\bar y\approx1.42\%$ on the training set), so each head is trained with class-weighted binary cross-entropy (positive weight $\pi=(1-\bar y)/\bar y\approx69$) computed only on frames where the listener is silent; the total objective is $\mathcal{L}=\mathcal{L}_{\mathrm{bc}}+\mathcal{L}_{\mathrm{st}}$ (Appendix~\ref{app:loss}). The \textsc{start} head does not feed the \textsc{bc} head and serves purely as a control. The \textsc{start} head does not feed into the \textsc{bc} head and therefore serves purely as a control. We train with AdamW \citep{loshchilov2019adamw}, using batch size 16 and precomputed CPC features. Training is early-stopped on validation \textsc{bc} AUPRC with patience 6. We use seeds $\{13,21,42\}$ for the main experiments and seed 13 for the remedy matrix in Section~\ref{sec:remedies}. 

\paragraph{Remedies.} 
We test three optional components aimed at the seen/unseen gap identified in Section~\ref{sec:conditioning}, with each component corresponding to a different hypothesis. 

\emph{Low-rank adapter} (memorization hypothesis). The fully fine-tuned trunk reaches its best validation AUPRC within the first few epochs and does not improve with further training (Appendix~\ref{app:training}). One possibility is that the model has more capacity than the transferable signal requires and uses the excess capacity to fit listener-specific patterns. If so, restricting the number of trainable parameters should encourage the model to rely more on transferable cues. We therefore freeze the trunk and insert rank-8 bottleneck adapters \citep{houlsby2019adapters,hu2022lora} of the form $\mathbb{R}^{256}\!\to\!\mathbb{R}^{8}\!\to\!\mathbb{R}^{256}$ into the channel and fused representations. The up-projection is zero-initialized and added residually. This leaves 356K trainable parameters, an 11-fold reduction relative to full trunk training. 

\emph{Listener-identity adversary} (identity-shortcut hypothesis). If the cross-listener gap arises because the shared representation encodes listener identity and the model relies on it as a shortcut, making identity harder to recover should improve transfer. We train a DANN-style classifier \citep{ganin2016dann} over the 146 training listeners plus \textsc{unk}, using the time-pooled listener-stream representation and a gradient-reversal layer whose strength $\lambda$ increases from 0 to $\lambda_{\max}$ during training. Task and adversarial gradients are clipped separately at norm 1. The adversary is attached before cross-channel fusion and FiLM, which prevents it from satisfying its objective simply by suppressing the style vector (Appendix~\ref{app:dann}). 

\emph{Lexical stream} (missing-linguistics hypothesis). Perturbation studies of dyadic VAP models suggest that they rely more strongly on linguistic content than on prosody \citep{inoue2025yeah,ekstedt2022prosody}. Words may also transfer across speakers more readily than voice characteristics. To test whether missing lexical information explains the gap, we add a frame-aligned 32-d embedding of the \emph{other} participants' gold transcripts to channel~2 only, never to the listener stream. This is an oracle, non-causal signal that would not be available directly at inference time, making the test intentionally favorable. If even gold lexical information does not help, a streaming-ASR version is unlikely to do so, although a richer lexical representation could still behave differently.

\section{Results and discussion}
\label{sec:results}
We first localize the source of the zero-shot failure (Section~\ref{sec:localization}). We then show that retraining reveals a cross-listener gap that conditioning does not close (Section~\ref{sec:conditioning}) and that persists across every remedy we test, including adaptation using the listener's own data (Secs.~\ref{sec:remedies} to \ref{sec:adaptation}). Finally, we examine the dissociation that helps explain this pattern (Section~\ref{sec:dissociation}); our choice of metric relative to prior dyadic work is discussed in Appendix~\ref{app:f1}.

\subsection{The dyadic head fails, but the features do not}
\label{sec:localization}

Applied zero-shot, the dyadic backchannel head performs at chance, with AUROC $0.499$ overall, $0.481$ for seen listeners, and $0.512$ for unseen listeners. At first, this might suggest that the acoustic front-end itself does not transfer. However, a linear probe trained on the same frozen CPC features reaches $0.704$ on held-out frames, while retraining only the trunk raises performance to $0.751$ ($0.742$ without conditioning; Figure~\ref{fig:localization}). The encoder therefore retains useful information, but the head learned on clean two-speaker audio can no longer exploit it when channel~2 contains a mixture of speakers. The main bottleneck is thus the predictor rather than the representation. We also ruled out label-timing misalignment, checkpoint-loading errors, insufficient model capacity, and a mismatch between the training loss and evaluation metric (Appendix~\ref{app:eliminations}).

\subsection{Conditioning helps seen listeners, not unseen ones}
\label{sec:conditioning}

\begin{figure}[!t]
\centering

\begin{minipage}[t]{0.44\linewidth}
    \centering
    \includegraphics[width=\linewidth]{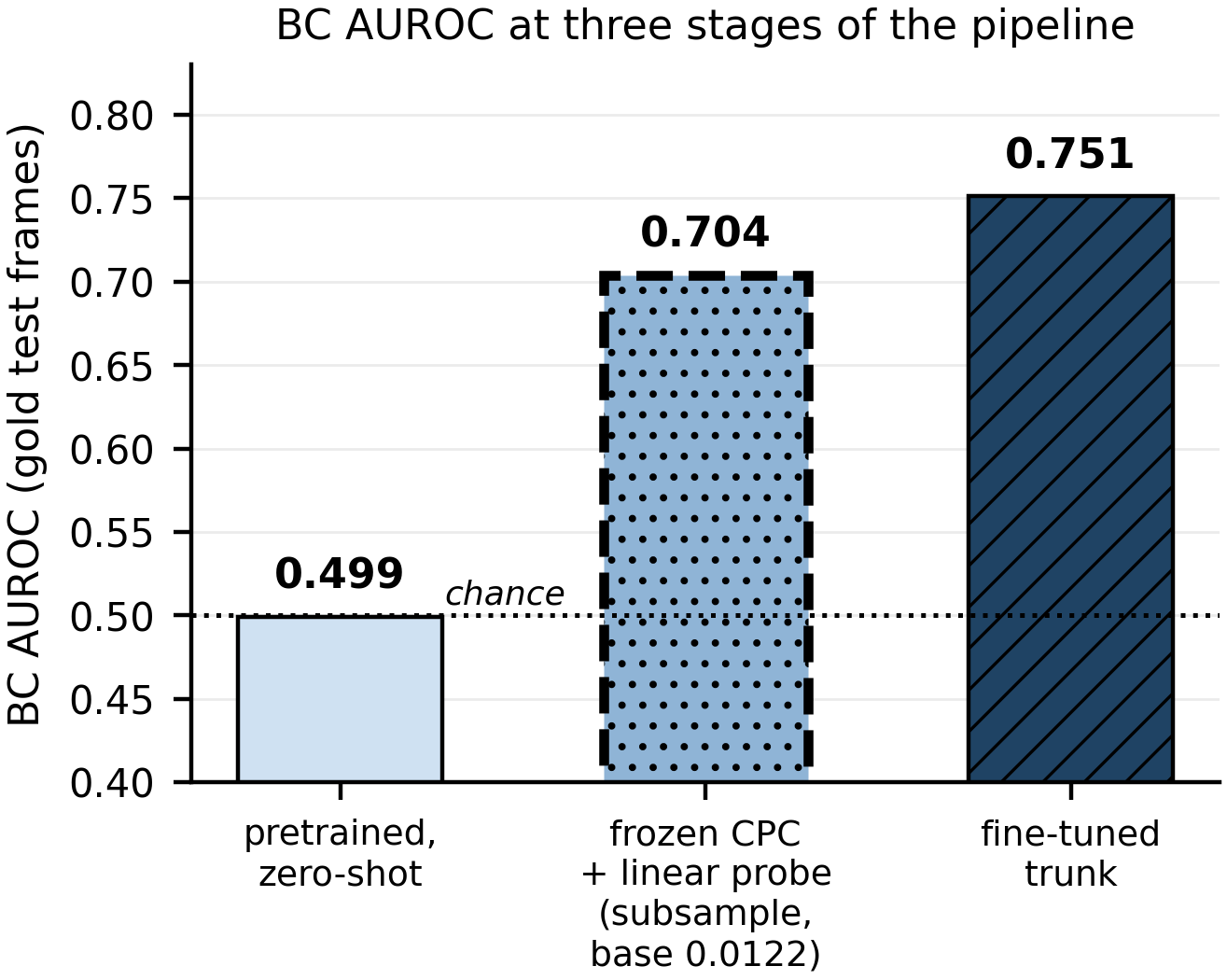}
    \caption{Backchannel AUROC across pipeline stages (independent measurements; the probe uses a separate subset).}
    \label{fig:localization}
\end{minipage}
\hfill
\begin{minipage}[t]{0.44\linewidth}
    \centering
    \includegraphics[width=\linewidth]{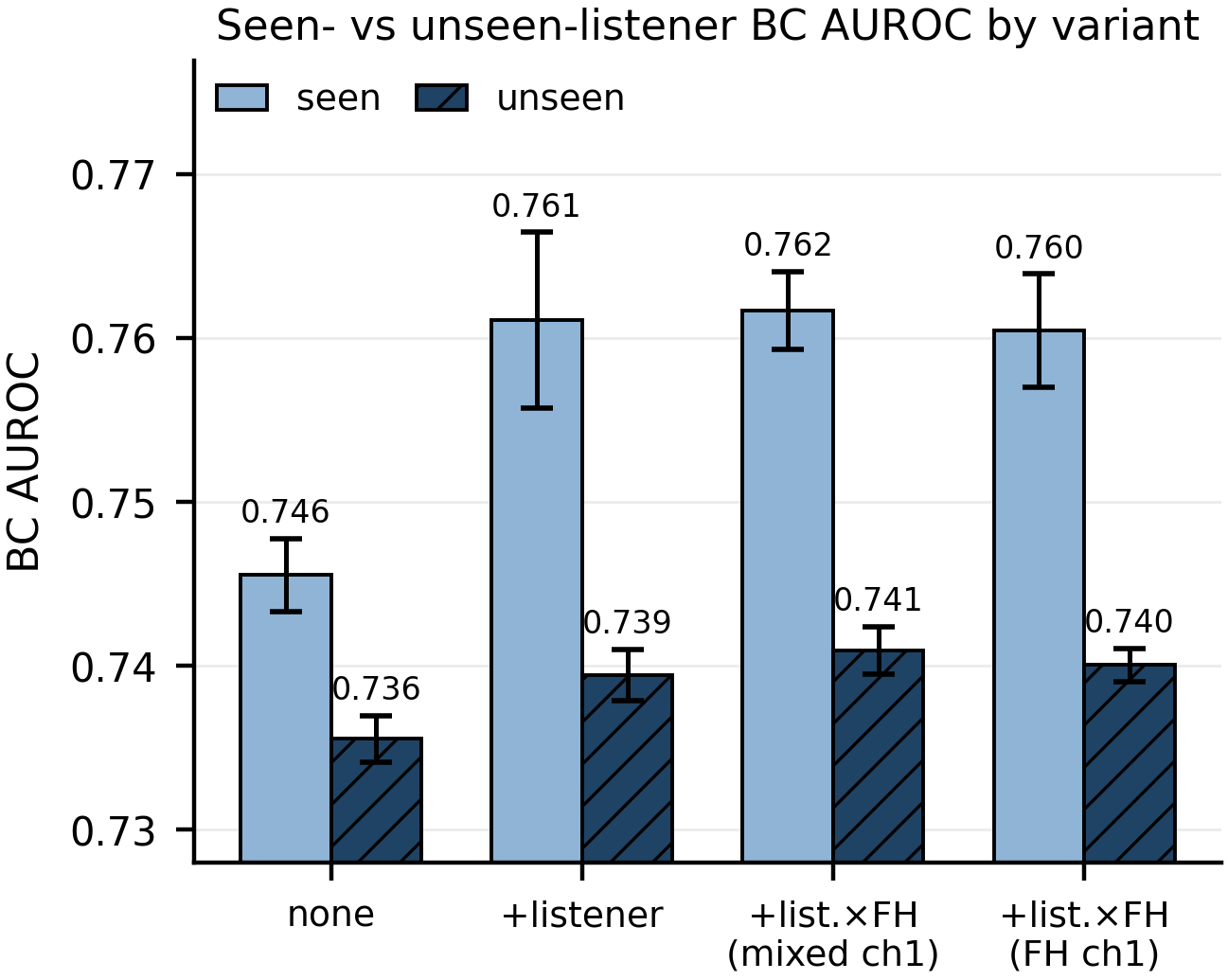}
    \caption{Backchannel AUROC for seen and unseen listeners by conditioning. (b)~Conditioning gain per slice: above zero for seen, not for unseen.}
    \label{fig:seen-unseen}
\end{minipage}

\end{figure}

Listener conditioning improves AUROC for seen listeners by $+0.016$ ($95\%$ CI $[+0.005,+0.025]$), but the improvement for unseen listeners is only $+0.005$ ($[-0.001,+0.011]$), with an interval that includes zero (Figure~\ref{fig:seen-unseen}). The corresponding $+0.011$ increase in the seen/unseen gap is not significant under the meeting-permutation null ($p=0.11$). We therefore report the two slices separately rather than interpret this as a significant interaction: conditioning clearly benefits listeners already seen during training, while any benefit for unseen listeners appears small at most. Event-level F1 shows the same within-slice pattern (Appendix~\ref{app:event_f1}). Conditioning increases seen-listener F1 from $0.064$ to $0.079$, while unseen-listener F1 remains within the variation across seeds. Adding floor-holder information, either as a label or as isolated audio, provides no benefit for unseen listeners. This is consistent with the behavioral result in Section~\ref{sec:dissociation}, where backchannel rate does not vary with the identity of the current floor-holder.

\subsection{The cross-listener gap resists every remedy}
\label{sec:remedies}

\begin{figure}[b]
\centering
\begin{minipage}[t]{0.44\linewidth}
  \centering
  \includegraphics[width=\linewidth]{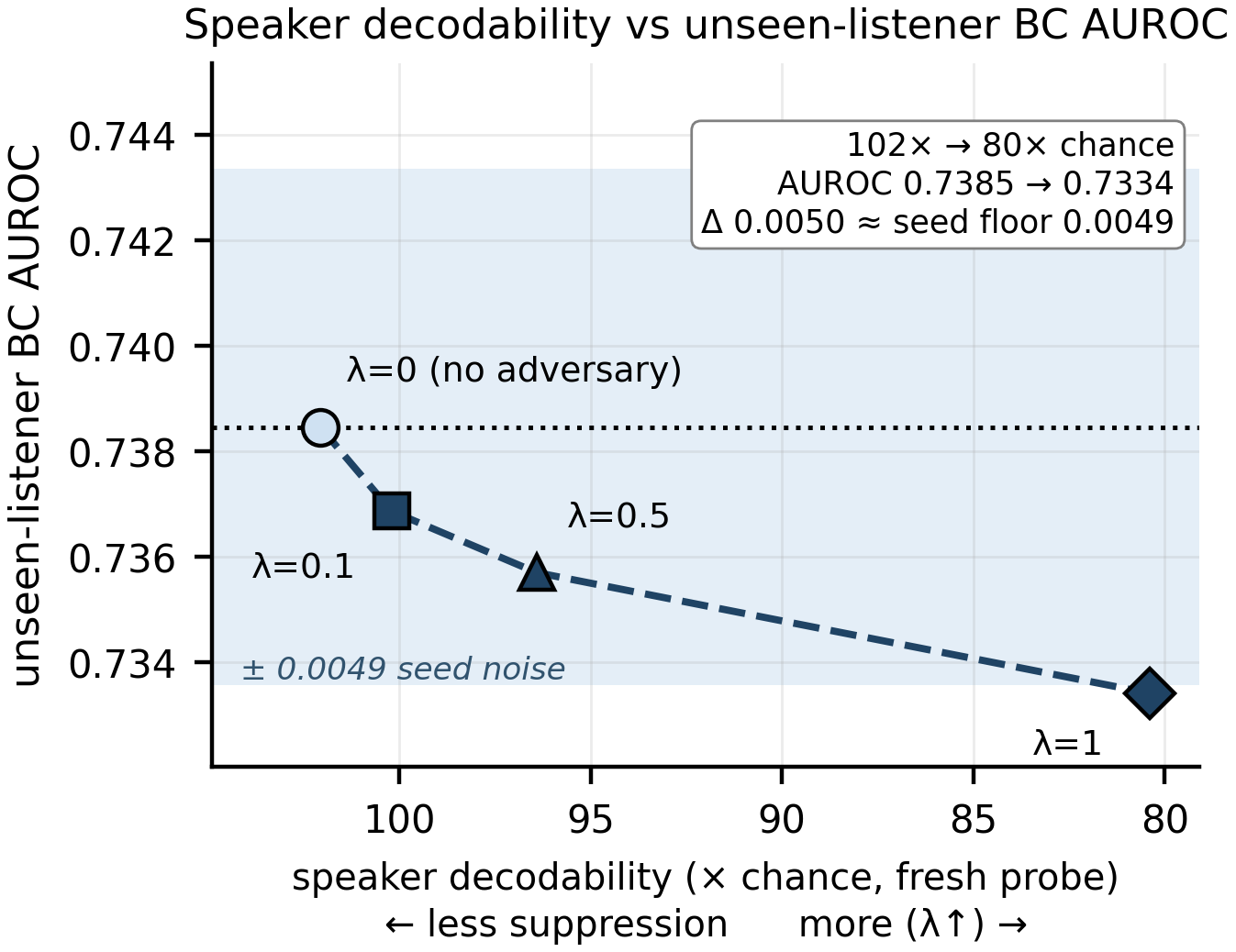}
  \caption{Unseen-listener backchannel AUROC vs.\ speaker-identity decodability (fresh probe, multiples of chance) across adversarial strengths $\lambda$. The y-range is narrow ($0.733$--$0.739$); the trend is the result.}
  \label{fig:dann}
\end{minipage}\hfill
\begin{minipage}[t]{0.44\linewidth}
  \centering
  \includegraphics[width=\linewidth]{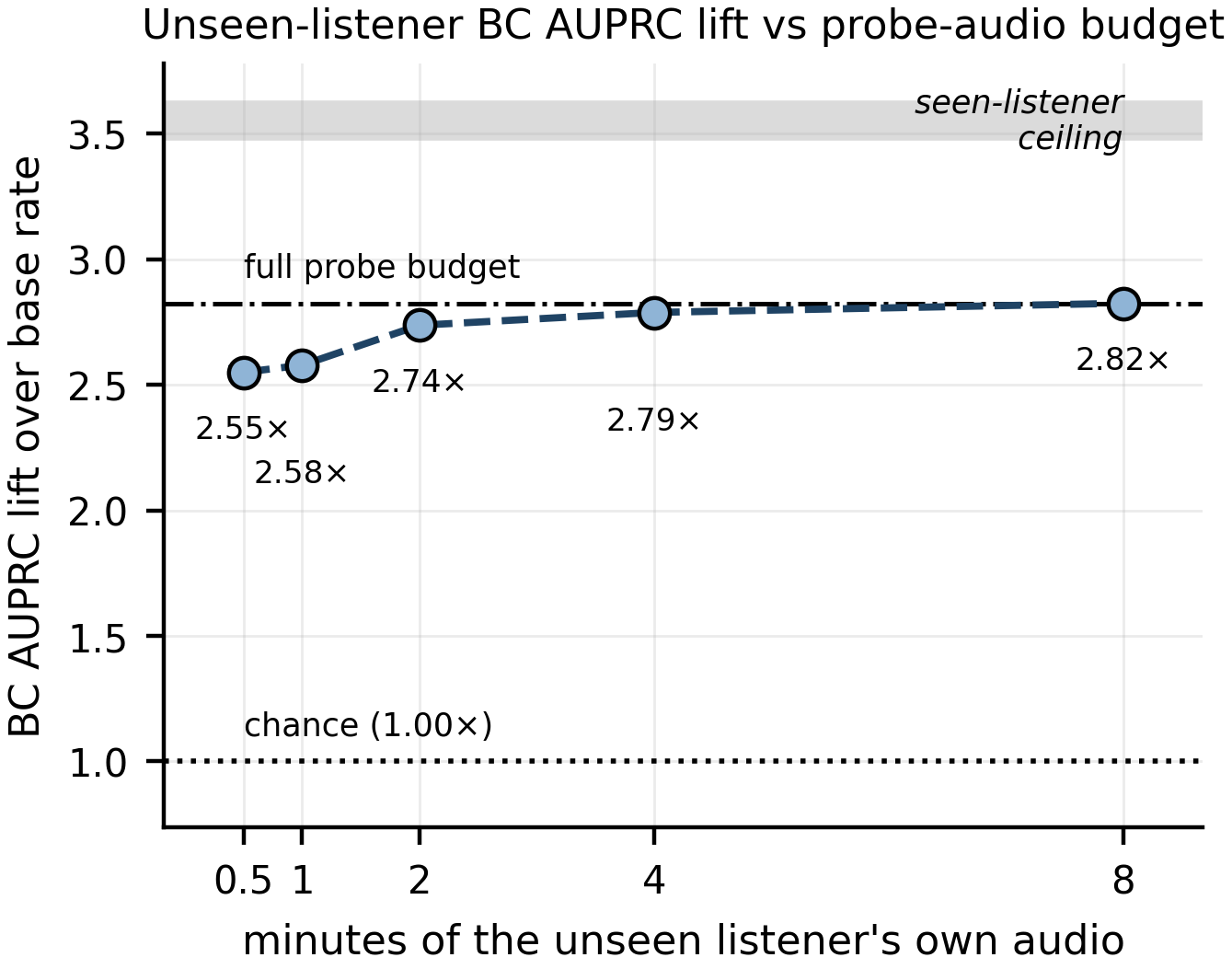}
  \caption{Unseen-listener backchannel AUPRC lift over the slice base rate vs.\ minutes of listener-specific adaptation audio. Band: seen-listener lift.}
  \label{fig:adaptation}
\end{minipage}
\end{figure}

If the model relies on speaker-specific acoustics rather than cues that generalize across people, then reducing memorization or suppressing identity should improve performance for unseen listeners. Table~\ref{tab:unseen_listener_remedies} shows that none of the remedies, either alone or in combination, produces a clear improvement. Every change falls within the three-seed spread ($0.0049$), so we do not interpret the small gains or losses. Two results are still informative. First, oracle lexical conditioning, which gives the model perfect non-causal access to the floor-holder's words, changes AUROC by only $+0.0025$. This is notable because perturbation studies of dyadic VAP models identify lexical content as an important cue \citep{inoue2025yeah,ekstedt2022prosody}. In the multi-party setting, lexical information therefore adds little to backchannel timing beyond what is already present in the acoustics. Second, adversarial training does not remove identity: a fresh probe still decodes it at $80\times$ chance even at $\lambda=1$, and stronger suppression consistently worsens backchannel prediction (Figure~\ref{fig:dann}, Appendix~\ref{app:dann}). Speaker-related acoustics are therefore not a removable shortcut but carry cues that support prediction, so identity and backchannel-relevant information are entangled in the representation.

\begin{table}[t]
\centering
\small
\caption{Remedies for unseen-listener generalization (seed 13); every change is within the three-seed spread ($0.0049$).}
\label{tab:unseen_listener_remedies}
\begin{tabular}{lcc}
\toprule
Method (listener\,+\,floor-holder base) & Unseen AUROC & $\Delta$ \\
\midrule
Baseline & 0.7385 & --- \\
+ Adapter ($r{=}8$) & 0.7307 & $-0.0078$ \\
+ Adversarial ($\lambda{=}0.1$) & 0.7369 & $-0.0016$ \\
+ Oracle lexical & 0.7409 & $+0.0025$ \\
+ Adapter\,+\,adversarial & 0.7302 & $-0.0083$ \\
+ Adapter\,+\,adv.\,+\,lexical & 0.7367 & $-0.0018$ \\
\bottomrule
\end{tabular}
\end{table}

\subsection{Adaptation does not close the gap either}
\label{sec:adaptation}

A final possibility is that unseen listeners simply require some of their own data. Adapting the model on up to eight minutes of audio from each held-out listener increases unseen-listener AUPRC only from $0.030$ to $0.033$, corresponding to a lift over the unseen base rate from $2.55\times$ to $2.88\times$, and most of the improvement occurs within the first two minutes (Figure~\ref{fig:adaptation}). For comparison, seen-listener lift ranges from $3.2$ to $3.6\times$ across configurations. The two slices differ in base rate by $1.45\times$, and because AUPRC does not scale proportionally with prevalence, lift only partially corrects for this difference (Appendix~\ref{app:event_f1}). We therefore focus on the shape of the adaptation curve rather than the absolute gap between slices. If limited listener-specific data were the main bottleneck, performance should continue to improve as more adaptation audio is added. Instead, the curve saturates after roughly two minutes and remains below the seen-listener range even after this partial base-rate correction. Together with Section~\ref{sec:remedies}, this argues against both candidate explanations: identity does not appear to be a removable shortcut, since suppressing it does not improve transfer, and the problem is not simply a lack of listener-specific data, since providing more of it brings little additional benefit.

\subsection{Backchannel is fragile across listeners; turn-onset is not}
\label{sec:dissociation}

\begin{figure}[!t]
\centering

\begin{minipage}[t]{0.44\linewidth}
    \centering
    \includegraphics[width=\linewidth]{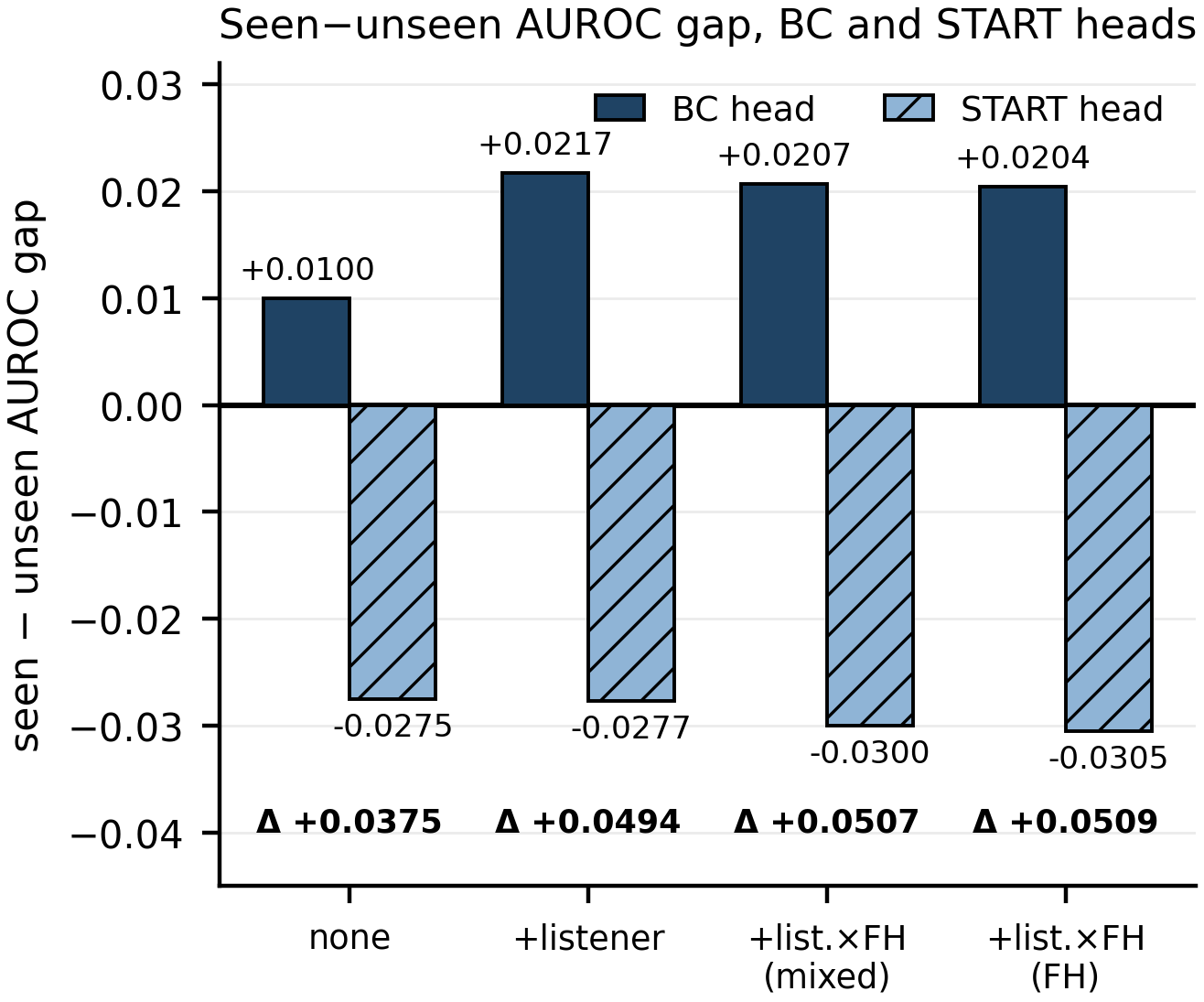}
    \caption{Seen$-$unseen AUROC gap for the backchannel and turn-onset heads, by variant. The difference between gaps ($\Delta=+0.038$ to $+0.051$) is the robust quantity; each gap alone is confounded by meeting sampling (Section~\ref{sec:eval}).}
    \label{fig:dissociation}
\end{minipage}
\hfill
\begin{minipage}[t]{0.44\linewidth}
    \centering
    \includegraphics[width=\linewidth]{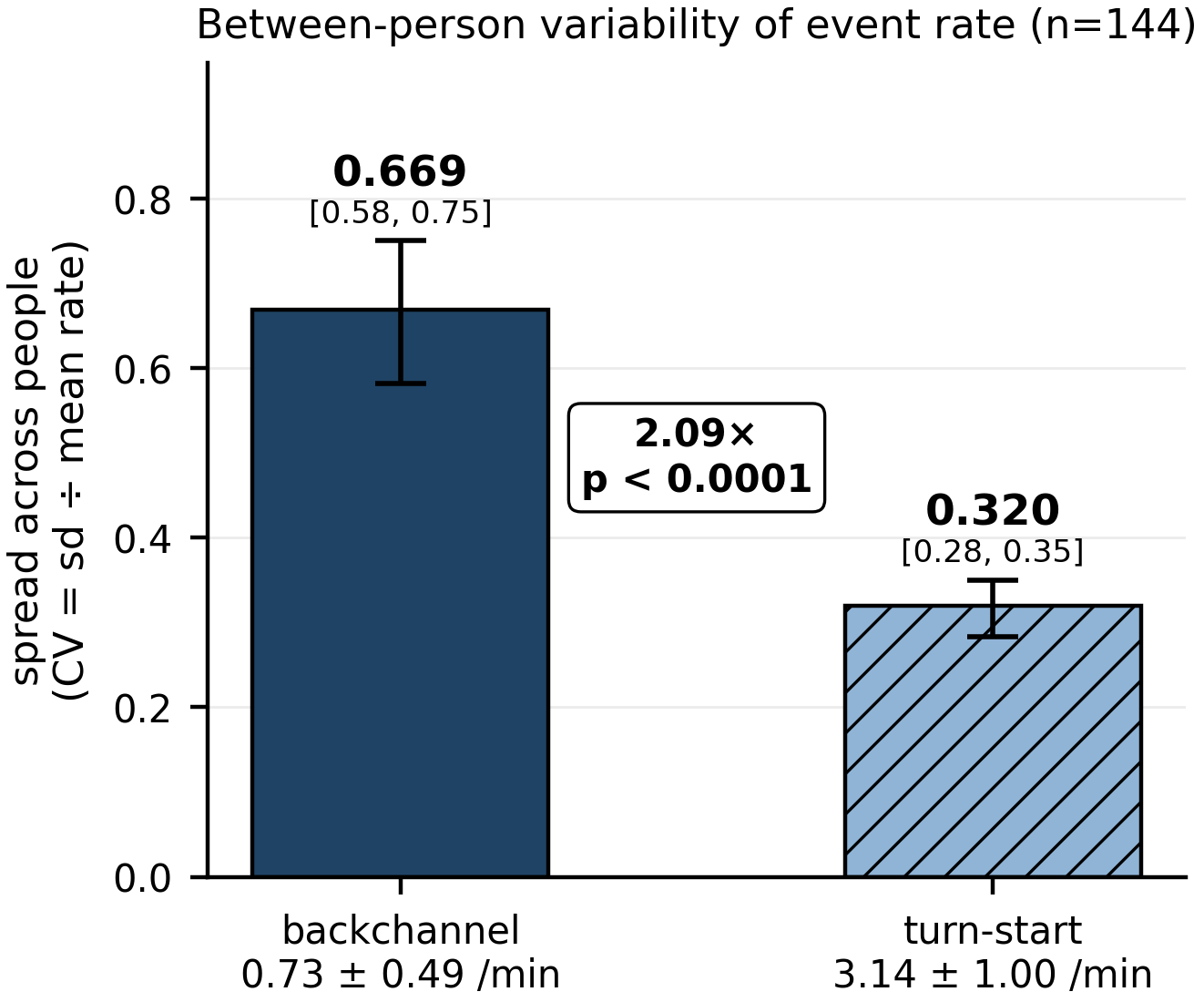}
    \caption{Between-person coefficient of variation of event rate ($n=144$, bootstrap $95\%$ CIs): backchannel $0.669$ vs.\ turn-onset $0.319$ ($p<0.0001$).}
    \label{fig:dispersion}
\end{minipage}

\end{figure}

The failures described in Secs.~\ref{sec:conditioning} to \ref{sec:adaptation} are specific to backchanneling. Using the same trunk, acoustic features, training data, listeners, and splits, turn-onset prediction behaves differently across the listener split (Figure~\ref{fig:dissociation}). Under the meeting-permutation null, neither gap is significant on its own: the backchannel gap is $+0.018$ ($p=0.41$), while the turn-onset gap is $-0.029$ ($p=0.09$). The latter appears to reflect the slower pace of the held-out meetings rather than better generalization, since per-meeting turn-onset AUROC is strongly related to turn-onset density (Spearman $\rho=-0.80$). What remains significant is the difference between the two heads, $\mathrm{BC}-\mathrm{START}=+0.047$, which is $2.8\times$ the standard deviation of the null distribution ($p=0.003$). 

Because both heads are evaluated on exactly the same frames, variation due to meeting composition is shared and largely cancels, leaving the behavior-specific difference. The two tasks are similarly learnable for seen listeners, with turn-onset AUROC between $0.757$ and $0.761$ and backchannel AUROC between $0.746$ and $0.762$. What differs is how well they transfer to new people. This contrast between the two heads, rather than either gap considered alone, is the main result. A similar dissociation appears in the behavior itself. 

Across the 144 participants with gold annotations, the between-person coefficient of variation is $0.669$ $[0.582,0.750]$ for backchannel rate and $0.319$ $[0.283,0.350]$ for turn-onset rate. The difference is $+0.350$ $[0.261,0.436]$ ($p<0.0001$; Figure~\ref{fig:dispersion}), meaning that backchanneling varies about $2.1\times$ more across individuals than turn-taking. At the same time, backchannel rate does \emph{not} meaningfully depend on who holds the floor. Only $0.034$ $[0.016,0.068]$ of listeners show variation across floor-holders beyond the within-meeting null, which is indistinguishable from the $0.05$ expected by chance. Backchannel behavior therefore depends much more on who the listener is than on who is currently speaking to them, consistent with the finding in Section~\ref{sec:conditioning} that floor-holder information provides no additional benefit. Taken together, these results suggest a coherent explanation. Turn-taking onset is a relatively structural behavior that can be learned from one set of people and transferred to others. Backchanneling varies much more from person to person, leaving a smaller listener-agnostic acoustic component for the model to exploit. This helps explain why conditioning improves prediction for seen but not unseen listeners, why suppressing or adding identity information does not improve transfer, why listener-specific adaptation quickly plateaus, and why even oracle lexical information provides little benefit. The limiting factor therefore appears to lie in the person-specific nature of the behavior itself rather than simply in the capacity or design of the model.

\section{Conclusion and perspectives} \label{sec:conclusion} 

A dyadic backchannel predictor applied directly to meetings performs at chance, but the failure comes from the predictor rather than the acoustic representation. The frozen encoder still linearly captures backchannel-relevant structure (AUROC $0.704$), and retraining the predictor recovers performance to $0.751$. This recovery reveals a more difficult limitation: a gap between seen and unseen listeners that remains under listener conditioning, capacity reduction, adversarial identity removal, per-listener adaptation, and oracle lexical conditioning. Under the same features, training data, and splits, no comparable gap appears for turn-onset prediction ($\mathrm{BC}-\mathrm{START}=+0.047$, $p=0.003$). Backchanneling also varies about twice as much across individuals as turn-taking and depends more on who the listener is than on who is speaking to them. The listener-agnostic acoustic signal available for prediction therefore appears limited. 

Two methodological lessons emerge. First, backchannels are rare, and multi-party conversation does not concentrate them as in dyads. Base-rate-independent ranking and listener-active event metrics are therefore more reliable than frame-level F1. The event-level seen/unseen gap is significant under the meeting-permutation null but disappears when false-alarm rates are matched, because prevalence differs by $1.5\times$ (Appendix~\ref{app:event_f1}). Second, listener-disjoint splits are also meeting-disjoint, so seen/unseen effects should be tested against meeting variation rather than seed variance; doing so changes which effects remain significant.

The direction we have not explored is architectural. Every model here combines the other participants into one channel, even though the dyadic head was not designed for mixtures. Per-speaker representations with floor-holder routing are a natural next step, together with signals unavailable to us, including a \emph{causal} lexical stream from streaming ASR, gaze, gesture, and visual grounding. Finally, if annotators agree only weakly on backchannel \emph{timing}, part of the gap may be aleatoric rather than epistemic. The released benchmark is intended to help quantify this human ceiling.


\bibliographystyle{plainnat}
\bibliography{references}


\appendix \providecommand{\TODO}[1]{}

\section{AUROC versus frame-level F1}
\label{app:f1}
We rank systems by AUROC and report frame-level F1 only for comparison with prior work. Across our 19 runs the two correlate (Spearman $\rho=0.82$) but do not always rank individual runs in the same order: the seed-21 \texttt{none\_mixed} run is fifth by validation-tuned F1 and fifteenth by AUROC. At a $1.65\%$ positive rate the F1-optimal threshold is set by the top ${\sim}2\%$ of frames, whereas AUROC scores the whole ranking, so a model can do well in the extreme upper tail while ranking the rest poorly. F1 spans only $0.024$ across all 19 runs, so we do not interpret small F1 differences between models.

For external context, our validation-tuned listener-active F1 is $0.077$--$0.101$, whereas \citet{inoue2025yeah} report $42.85$ (precision $32.5$, recall $62.8$) on Japanese Wizard-of-Oz attentive-listening dialogues with trained operators as listeners, at a ${\sim}10\%$ positive rate, about six times ours. Our 1-s labeling horizon (versus their 500~ms) would if anything raise our rate, and no listener-active slice of meetings approaches dyadic density (Appendix~\ref{app:selectivity}); the difference reflects curated dyadic data versus natural meetings, not model quality.

\section{Benchmark construction details} \label{app:benchmark} \paragraph{Rule-based backchannel labels.} Of the 171 meetings, 139 contain gold \texttt{bck} dialogue-act annotations; the remaining 32 are labeled using a rule-based procedure. A candidate is a listener utterance lasting at most 1.2~s and containing at most three tokens, whose transcript matches a 16-entry lexicon (mm-hm, yeah, okay, right, \dots), with another participant speaking within 0.5~s before and 1.5~s after it. Listener speech regions are first merged when separated by gaps shorter than 0.25~s, and event durations are capped at 3.0~s. The rule does not filter based on preceding listener speech. When applied to the 139 gold meetings (556 views), it proposes 14{,}498 events against 12{,}594 gold events and recovers them with event-level precision $0.634$, recall $0.729$, and F1 $0.678$ using one-to-one greedy onset matching with a $\pm300$~ms collar and exact-millisecond timing. Rule-labeled segments are used only for training, accounting for 19{,}236 segments or $40\%$ of the training set. Every metric reported in the paper is computed on gold-annotated meetings. \paragraph{Turn-onset labels.} A \textsc{start} event is the onset of a listener speech region that is not a backchannel. Since regions separated by gaps shorter than 0.25~s are merged, each \textsc{start} is preceded by more than 0.25~s of listener silence. \textsc{start} targets use the same future shift, $H=1$~s, as the backchannel targets and are evaluated on the same listener-silent frames. \paragraph{Split composition.} Table~\ref{tab:splits} summarizes the dataset splits. Training, validation, and test \emph{meetings} are disjoint, and validation listeners are also disjoint from all other listeners. The unseen test portion contains six complete AMI participant groups, corresponding to 24 people who appear in neither the training nor validation set. The seen test portion contains 14 additional meetings involving 56 participants who also appear in the training set, but in different meetings. This overlap makes the seen arm a personalization test and explains why the listener counts do not sum to 190. The two test slices also differ in backchannel prevalence: $2.06\%$ of listener-active frames are positive for seen listeners, compared with $1.41\%$ for unseen listeners. This corresponds to $1.48$ versus $0.98$ events per listener-active minute, a $1.5\times$ difference that is itself significant under the meeting-permutation null of Section~\ref{sec:eval} ($p=0.002$). Metrics that depend on the base rate, such as F1 and AUPRC, are therefore comparable across configurations within a slice but not directly across the two slices. Cross-slice comparisons in the paper instead rely on AUROC (Appendix~\ref{app:event_f1}). \begin{table}[htbp] \centering \small \caption{Masked-listener benchmark by split. Views are meeting-listener pairs; segments are 20-s windows. Listener counts do not sum to 190 because all 56 seen-test listeners are also training listeners. Backchannel events are attributed to their view; 19 of the 18{,}697 occur after the last complete 20-s segment and therefore appear in no segment.} \label{tab:splits} \begin{tabular}{lrrrrr} \toprule Split & Meetings & Views & Listeners & Segments & BC events \\ \midrule Train & 113 & 450 & 146 & 47{,}780 & 13{,}626 \\ Validation & 20 & 80 & 20 & 7{,}744 & 1{,}574 \\ Test, seen & 14 & 56 & 56 & 6{,}044 & 1{,}655 \\ Test, unseen & 24 & 96 & 24 & 10{,}020 & 1{,}842 \\ \midrule Total & 171 & 682 & 190 & 71{,}588 & 18{,}697 \\ \bottomrule \end{tabular} \end{table} \section{Eliminated explanations for the zero-shot failure} \label{app:eliminations} Section~\ref{sec:localization} attributes the chance-level zero-shot AUROC ($0.499$) of the dyadic head to the predictor rather than the representation. Before drawing that conclusion, we ruled out four simpler explanations. \paragraph{Label-timing misalignment.} One possibility is that our future-shifted targets use a different timing convention from the one used to train the pretrained head. Correct predictions could then fall on the wrong frames and appear uninformative. We re-scored the zero-shot head against targets shifted from $-1.0$ to $+1.0$~s in $0.1$-s steps, for a total of 21 offsets. AUROC ranges from $0.341$ to $0.527$, with the maximum at $+0.4$~s and a value of $0.499$ at zero offset. No shift recovers a meaningful signal, indicating that the head is uninformative on meeting audio rather than simply misaligned in time. \paragraph{Checkpoint loading.} An incorrectly loaded checkpoint, for example because of missing keys, an incorrect module mapping, or a re-initialized head, could also produce chance-level predictions. Strict state-dict loading reports zero missing keys. The only unexpected key is a VAP codebook embedding that is not used by any of our prediction heads, while the 28 encoder tensors are duplicated into the two tied channel encoders by design. More importantly, the same loaded encoder supports the $0.704$ linear probe and the $0.742$ retrained trunk reported in Section~\ref{sec:localization}, which would not be possible if the weights were corrupted. The checkpoint is therefore intact; what changes is the input distribution. \paragraph{Capacity.} The head could also fail because it is too small to express the mapping required for multi-party audio. If this were the case, however, the same architecture should remain limited after retraining. Instead, it reaches AUROC $0.742$ in the unconditioned setting and $0.751$ with conditioning once the trunk is retrained (Section~\ref{sec:localization}). Capacity therefore does not explain the zero-shot failure. We did not additionally train a larger prediction head. \paragraph{Loss/metric mismatch.} Training uses class-weighted cross-entropy with early stopping on AUPRC, whereas systems are ranked by AUROC. The zero-shot head, however, is at chance under every criterion rather than AUROC alone: its AUPRC is $0.0140$ against a base rate of $0.0137$, corresponding to a lift of $1.02$. The training loss and evaluation metrics also move consistently. On a 4{,}000-segment training subset evaluated on held-out validation segments, validation loss decreases from $2.17$ to $1.18$ while validation AUPRC increases from $0.0136$ to $0.0343$. Across the 19 full runs, frame-level F1 and AUROC have a Spearman rank correlation of $\rho=0.82$ (Appendix~\ref{app:f1}). We therefore find no evidence that the zero-shot result is caused by a mismatch between the training objective and evaluation metric. We did not additionally train with an AUROC surrogate. 

\section{Training objective}
\label{app:loss}
Each head is trained with class-weighted binary cross-entropy on the listener-silent frames selected by a mask $m_t$ (the frames where a backchannel decision is live):
\begin{equation}
\mathcal{L}_{\mathrm{bc}}=-\frac{1}{\sum_t m_t}\sum_t m_t\Big[\pi\,y^{\mathrm{bc}}_t\log\hat p^{\mathrm{bc}}_t+(1-y^{\mathrm{bc}}_t)\log(1-\hat p^{\mathrm{bc}}_t)\Big],
\qquad \mathcal{L}=\mathcal{L}_{\mathrm{bc}}+\mathcal{L}_{\mathrm{st}},
\end{equation}
where $y^{\mathrm{bc}}_t=\mathbbm{1}[\exists\,\tau\in(t,t+H]:\ \tau\text{ is a backchannel onset}]$, $\hat p^{\mathrm{bc}}_t=\sigma(w_{\mathrm{bc}}^\top h'_t)$, and $\pi=(1-\bar y)/\bar y\approx69$ from the training base rate $\bar y\approx1.42\%$. $\mathcal{L}_{\mathrm{st}}$ is identical in form on the \textsc{start} targets. The unshifted onset times are stored separately for the event-level evaluation of Appendix~\ref{app:event_f1}.

\section{Training dynamics} 
\label{app:training} 

\begin{figure}[htbp] \centering \includegraphics[width=0.62\linewidth]{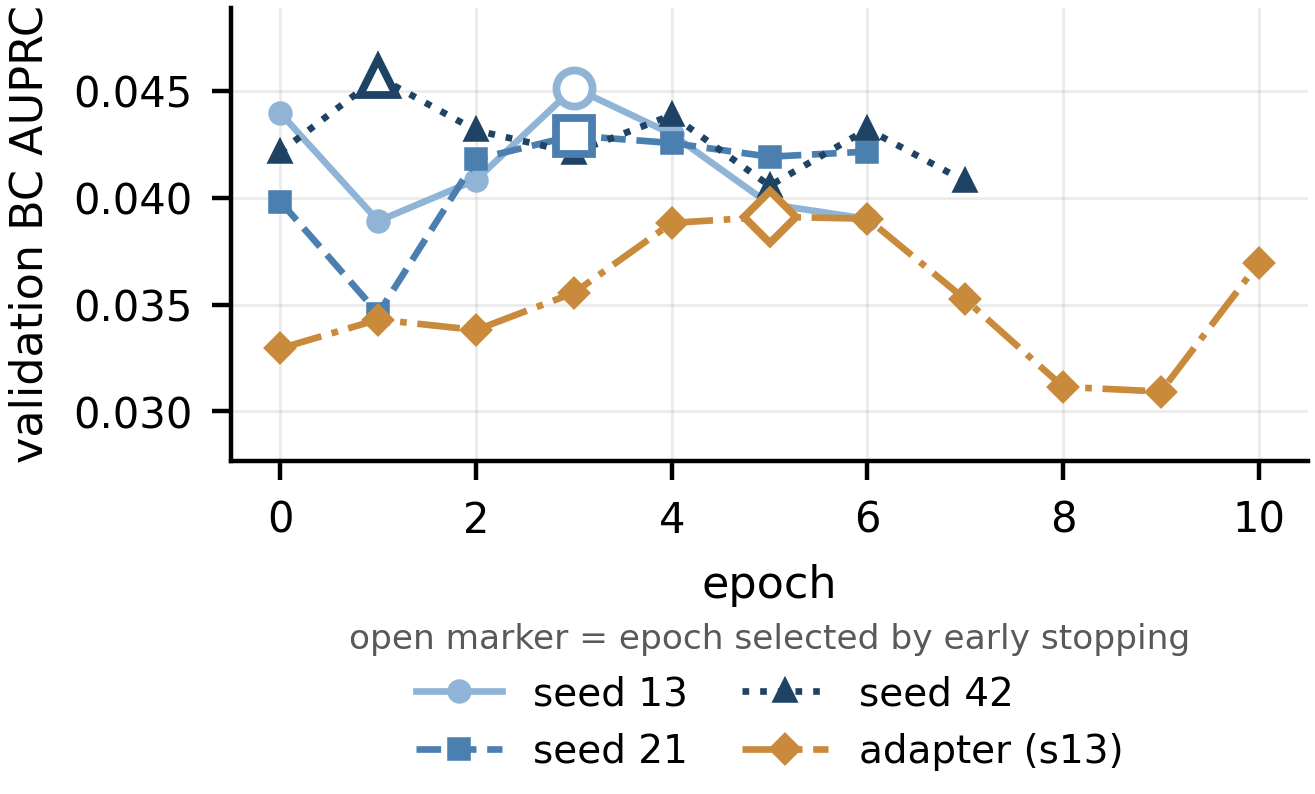} \caption{Validation backchannel AUPRC per epoch for the fully fine-tuned trunk across three seeds and the rank-8 adapter for seed 13. Epoch 0 is the first completed pass; open markers indicate the checkpoints selected by early stopping.} \label{fig:training} \end{figure}

Figure~\ref{fig:training} shows validation backchannel AUPRC across epochs for the fully fine-tuned trunk in the \emph{listener+floor-holder} configuration with mixed channel~2, using seeds 13, 21, and 42, together with the rank-8 adapter for seed 13. The full trunk reaches its best performance within the first four passes. Its peaks occur at epoch indices 3, 3, and 1, with AUPRC values of $0.0452$, $0.0429$, and $0.0457$, respectively. Performance does not improve afterward and instead fluctuates within a narrow range until patience stops training at epoch 6 or 7. Across the full grid of 19 runs, the best checkpoint occurs between epoch indices 0 and 5. Early stopping, with patience 6 and min-delta $0.002$ on validation AUPRC, selects the strict argmax checkpoint in all 19 runs. The two criteria can differ because the argmax is not min-delta gated, but in practice they never do. The adapter learns more slowly and reaches a lower peak. Its best checkpoint is epoch index 5 at AUPRC $0.0391$, below the peak reached by each fully fine-tuned trunk, and training continues to epoch 10 before patience is exhausted. This provides the training-side view of the adapter result in Table~\ref{tab:unseen_listener_remedies}: restricting the trunk to a rank-8 update reduces validation performance without improving generalization.  \section{Listener-adversarial training details} \label{app:dann} \paragraph{Classifier and pooling.} A classifier $g_\phi$ predicts listener identity from the listener-stream representation $a_1\in\mathbb{R}^{400\times256}$. When an adapter is present, $a_1$ is taken from its output; otherwise, it is taken from the output of the channel-wise transformer. The classifier has $L=147$ classes, corresponding to the 146 training listeners plus $\langle\mathrm{unk}\rangle$. No training example carries the unknown label, so 146 classes are reachable during training. Because identity is constant within each segment, we pool $a_1$ over time by concatenating the per-dimension mean and standard deviation, producing a 512-d vector. This is mapped by a linear layer to 256 units, followed by a GELU nonlinearity and dropout ($p=0.1$), then a second linear layer to the 147 logits; no normalization layer is used.

\paragraph{Objective and schedule.} The adversarial loss is the cross-entropy of $g_\phi$ over batch examples with a known listener, \begin{equation} \mathcal{L}_{\mathrm{adv}} = -\frac{1}{\lvert\mathcal{B}_{+}\rvert} \sum_{i\in\mathcal{B}_{+}} \log\Big[g_{\phi}\big(R_{\lambda}(a_{1}^{(i)})\big)\Big]_{\ell_i}, \qquad \lambda(p)=\lambda_{\max}\left(\frac{2}{1+\exp(-10p)}-1\right), \end{equation} where $\ell_i$ is the true listener label, $[\,\cdot\,]_{\ell_i}$ selects the corresponding predicted probability, and $\mathcal{B}_{+}$ excludes $\langle\mathrm{unk}\rangle$ examples. The gradient-reversal layer $R_\lambda$ \citep{ganin2016dann} behaves as the identity on the forward pass, allowing $g_\phi$ to learn listener identity, while multiplying the gradient by $-\lambda$ on the backward pass. This pushes the trunk toward a representation from which listener identity is harder to recover. The adversarial strength follows the standard DANN schedule as a function of the fraction of completed training $p$. The values of $\lambda$ reported in Table~\ref{tab:unseen_listener_remedies} and Figure~\ref{fig:dann} are $\lambda_{\max}$, the saturation value of the ramp rather than a constant strength, where $p=(e+b/B)/E$ is the fraction of planned training completed (clamped to $[0,1]$) for epoch index $e$, batch $b$ of $B$, and $E$ planned epochs. Because $\lambda$ is held at zero during evaluation, the adversary never perturbs a reported metric. 
We add $\mathcal{L}_{\mathrm{adv}}$ to the task objective from Section~\ref{sec:model}. Task and adversarial gradients are clipped separately at norm~1, so varying $\lambda$ changes the strength of the invariance pressure without directly changing the effective task update. \paragraph{Attach point.} The adversary reads $a_1$ before cross-channel fusion and FiLM. If it were attached at or after FiLM, the model could satisfy the adversarial objective simply by suppressing the conditioning vector $z$, which would remove the very signal the method is intended to make robust. Attaching it to $a_1$ instead allows the adversary to target incidental listener-acoustic identity while leaving the style vector intact. Using the listener stream rather than the fused representation also makes the objective specific to \emph{listener} identity, while the post-adapter placement ensures that reversed gradients reach trainable parameters when the trunk is frozen. 

\paragraph{Residual-identity probe.} The training-time accuracy of $g_\phi$ is not a reliable measure of how much listener identity remains in $a_1$. Under gradient reversal, the classifier itself can collapse toward chance even when the representation still contains recoverable identity information. We therefore measure residual identity with a \emph{fresh} probe trained after adversarial training on the frozen, pooled $a_1$. The probe reuses the adversary architecture unchanged, with Linear($512!\to!256$), GELU, dropout $0.1$, and Linear($256!\to!147$), but removes the gradient-reversal layer. It reads the same representation as the adversary: the listener channel after the channel-wise transformer and, when present, the adapter, pooled over time by concatenating its mean and standard deviation. The trunk remains frozen and only the probe head is trained. Training uses 43{,}002 training-split segments for up to 12 epochs at a learning rate of $10^{-3}$, with early stopping (patience 3) on a disjoint set of 4{,}778 segments from the same split, containing the same speakers but different segments. The probe is then evaluated once on the 6{,}044 held-out seen-listener test segments, the same seen-test portion used throughout the paper. Chance performance is $1/146\approx0.69\%$ top-1 accuracy. Table~\ref{tab:probe} reports the result for each adversarial strength, while Figure~\ref{fig:dann} expresses the same values as multiples of chance. Even without an adversary, listener identity is highly decodable at $69.9\%$ top-1 accuracy, or $102\times$ chance. The strongest adversary removes roughly one fifth of this accuracy, but still leaves $55.1\%$, corresponding to $80\times$ chance. 

\begin{table}[htbp] \centering \caption{Top-1 listener-identity accuracy of a fresh probe on the frozen listener-stream representation for each adversarial strength (146 classes; chance $0.69\%$).} \label{tab:probe} \begin{tabular}{lcc} \toprule $\lambda_{\max}$ & Top-1 accuracy & $\times$ chance \\ \midrule 0 (no adversary) & 69.9\% & 102.1 \\ 0.1 & 68.6\% & 100.2 \\ 0.5 & 66.0\% & 96.4 \\ 1.0 & 55.1\% & 80.4 \\ \bottomrule \end{tabular} \end{table}

\section{Selectivity of listener-active slices} \label{app:selectivity} 
\begin{figure}[htbp] \centering \includegraphics[width=0.55\linewidth]{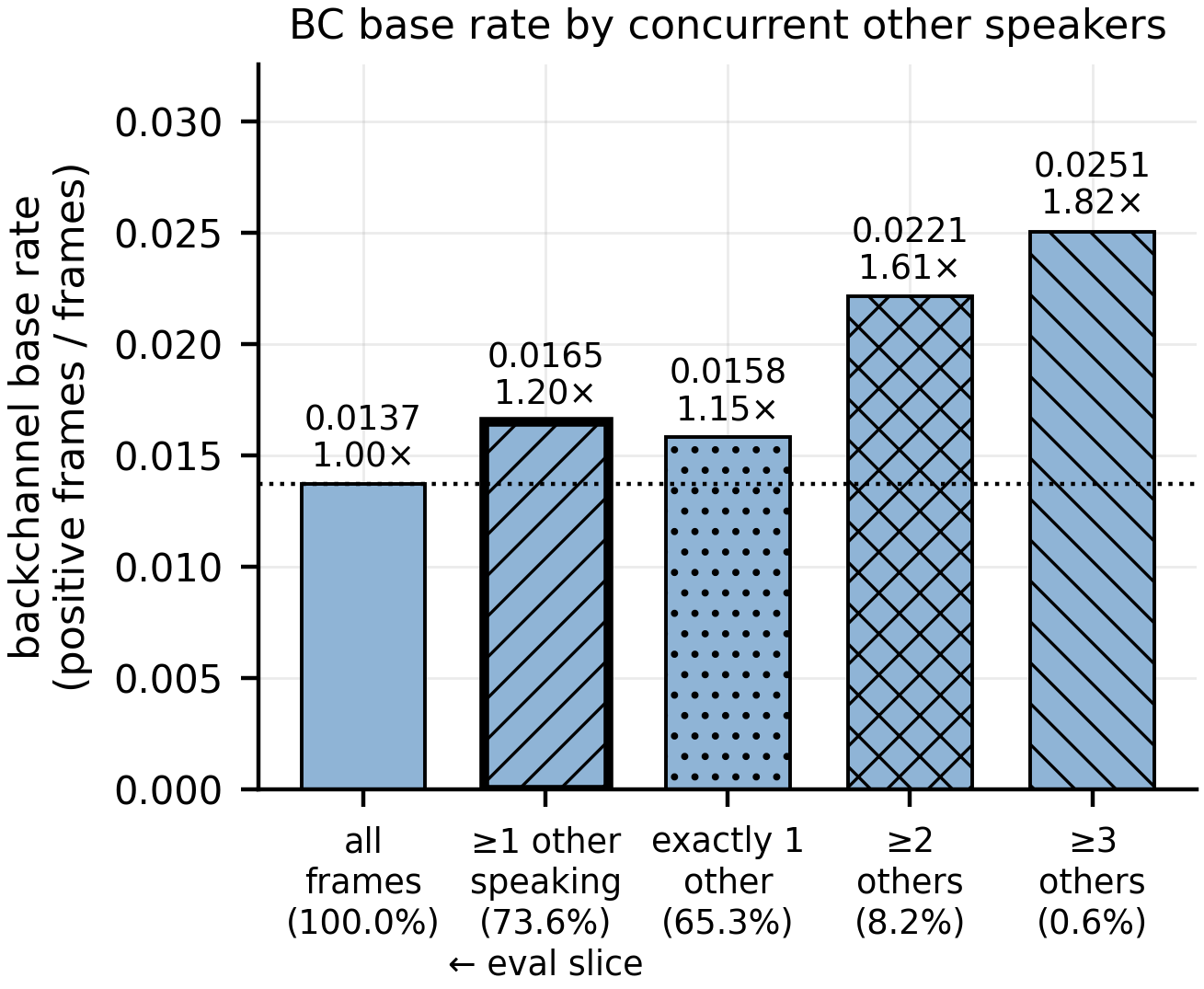} \caption{Backchannel base rate as a function of the number of concurrently active other speakers in the gold test set.} \label{fig:selectivity} \end{figure}

In dyadic conversation, scoring only frames in which the interlocutor is speaking removes many irrelevant frames and concentrates backchannels. In meetings, the same restriction is much less selective. Someone other than the listener is speaking $73.6\%$ of the time, so restricting evaluation to these frames increases the backchannel rate only from $1.37\%$ to $1.65\%$ of valid frames, a concentration factor of $1.20$. Figure~\ref{fig:selectivity} breaks this down by the number of concurrently active other speakers. The dominant case is exactly one active speaker, which accounts for $65.4\%$ of frames and has a positive rate close to the overall corpus average. Requiring two or more active speakers removes most of the data while producing only a small increase in density. No slice of multi-party data approaches the ${\sim}10\%$ density of curated dyadic sets. This explains why the absolute F1 values in Appendix~\ref{app:f1} cannot be made comparable to dyadic reports simply by choosing a more restrictive evaluation region.  \section{Event-level F1} \label{app:event_f1} Table~\ref{tab:event_f1} reports event-level precision, recall, and F1 for the four conditioning configurations from Section~\ref{sec:conditioning}, evaluated on the listener-active slice, where the listener is silent and at least one other participant is speaking. Events are defined as local maxima of the frame-level posterior above a threshold. We apply greedy non-maximum suppression with a minimum distance of 6 frames, equal to the collar at 20~Hz, and use no smoothing. A predicted event counts as a hit when it falls within $\pm300$~ms of an unshifted reference onset, with one-to-one matching. We report two timing conventions. Under \emph{onset}, the 1-s prediction horizon is left uncompensated, so a correctly fitted model that peaks early is penalized. Under \emph{window}, any peak from one prediction horizon before the onset through the collar after it counts as a hit. The threshold is chosen separately for each configuration and seed using the validation posteriors of the best checkpoint. We evaluate 19 values in $[0.05,0.95]$ and select the one maximizing onset event-F1, then apply it unchanged to the test set. The selected thresholds lie between $0.80$ and $0.90$, with $0.85$ chosen in 8 of the 12 runs. Validation event-F1 at these operating points ranges from $0.046$ to $0.074$, closely matching the test range. The result is not especially sensitive to threshold choice: F1 remains nearly flat between $0.80$ and $0.90$, the validation-selected threshold is also test-optimal in 68 of the 72 run$\times$slice$\times$convention cells, and the largest gain from tuning directly on test is only $0.005$ F1. Values are averaged over seeds 13, 21, and 42, with standard deviation reported for F1. Precision and recall are reported as means only because moving the threshold by one step can change recall by as much as $0.06$ while F1 remains on the same plateau. Within each slice, event-level F1 follows the same pattern as the AUROC analysis. Conditioning increases seen-listener F1 from $0.064$ to $0.079$ under the onset convention, while unseen-listener F1 remains within the variation across seeds at $0.041$ to $0.047$. Every configuration also gives higher F1 for seen than unseen listeners. As shown below, however, this difference is explained by the base rate rather than by personalization. Precision remains low throughout, between $0.03$ and $0.06$ under the onset convention. At a positive rate of only $1.65\%$, the models produce many false alarms, which is the event-level consequence of the base-rate issue discussed in the main text. \paragraph{The seen/unseen event-F1 gap is a base-rate effect.} Averaged across the 12 grid runs at their validation-selected thresholds, the event-F1 difference between seen and unseen listeners is $+0.029$ under the onset convention and $+0.043$ under the window convention. Because event-F1 is a pooled ratio, its meeting-permutation null is constructed from per-meeting (TP, FP, FN) counts, which are summed within each reassigned subset while leaving peak picking and matching unchanged. Pooling over all 38 test meetings reproduces the direct computation exactly. Under the size-matched 14-vs-24 null from Section~\ref{sec:eval}, the gap is significant for both conventions ($p=0.031$ onset, $p=0.009$ window; one test with a consistency check because both use the same peaks and threshold), whereas the AUROC gap is not ($p=0.41$; Section~\ref{sec:dissociation}). This does not indicate a personalization effect that AUROC fails to capture. The two slices differ in prevalence by $1.5\times$ (Appendix~\ref{app:benchmark}), and that prevalence difference is more significant under the same null than the F1 gap itself: $p=0.002$ for the positive-frame rate and $p=0.0008$ for events per minute. Since the precision term of F1 increases with prevalence even when ranking quality is unchanged, this difference alone can produce a higher F1. When prevalence is controlled, the gap disappears. We set the threshold separately for each slice so that both operate at the same false-alarm rate, using $1.55$, $2.90$, and $4.71$ false alarms per listener-active minute to span the selected operating points. Under this comparison, the seen/unseen recall gap ranges from $+0.016$ to $+0.040$ and is never significant ($p=0.22$ to $0.45$). Two prevalence-free evaluations, frame-level AUROC and recall at a matched false-alarm rate, therefore agree that seen and unseen listeners are not distinguishable beyond meeting sampling. The only metric that disagrees is the prevalence-dependent one, in a setting where prevalence itself differs. A 7-vs-7 null restricted to the seen meetings has roughly $1.7\times$ greater spread and likewise does not reach significance ($p=0.41$ / $0.31$). The permutation analysis captures meeting composition only; it does not include variation across runs or the once-per-run threshold-selection procedure. \paragraph{AUPRC lift.} The same qualification applies to AUPRC, whose chance level is the base rate: $0.0170$ for seen listeners and $0.0118$ for unseen listeners on valid frames. Expressed as lift relative to each slice's own base rate, the four grid configurations in the order shown in Table~\ref{tab:event_f1} achieve $3.17\times$, $3.55\times$, $3.53\times$, and $3.64\times$ for seen listeners, compared with $2.74\times$, $2.82\times$, $2.89\times$, and $2.79\times$ for unseen listeners. The adaptation curve in Section~\ref{sec:adaptation} increases from $2.55\times$ to $2.88\times$ as listener-specific audio grows from $0.5$ to $8$ minutes. Lift removes the leading effect of the base rate, but it does not make AUPRC fully prevalence-independent because AUPRC does not scale proportionally with prevalence. These values therefore narrow the cross-slice difference but do not resolve it. Within each slice, they reproduce the conditioning pattern reported in Section~\ref{sec:conditioning}. \paragraph{START at event level.} For completeness, we apply the same event-level evaluation to the turn-onset head. Its onset track is rebuilt from the labels and checked by reproducing the backchannel onsets on every validation and test frame before scoring. For \textsc{start}, seen and unseen event-F1 are $0.157$ and $0.147$, respectively, giving a gap of $+0.010$ with $p=0.28$. For \textsc{bc}, the corresponding values are $0.073$ and $0.044$. The resulting difference of gaps is therefore $\mathrm{BC}-\mathrm{START}=+0.019$ at event level ($p=0.29$), or $+0.036$ under the window convention ($p=0.11$). The frame-level AUROC dissociation reported in Section~\ref{sec:dissociation} ($+0.047$, $p=0.003$) therefore does not replicate as a difference between event-level F1 scores. This is expected. \textsc{start} events are $5.7\times$ more common than backchannels, with $19{,}887$ versus $3{,}493$ test onsets, so the two heads operate at different points on their precision-recall curves. A difference of F1 scores also lacks the shared scale and common chance level that make a difference in AUROCs a meaningful paired contrast. In addition, the \textsc{bc} term contains the prevalence artifact described above. For these reasons, the dissociation claim rests on the AUROC analysis, which provides the appropriate comparison. \begin{table}[htbp] \centering \small \setlength{\tabcolsep}{4pt} \caption{Event-level precision (P), recall (R), and F1 on the listener-active slice, using a $\pm300$~ms collar and a validation-selected threshold. Values are means over three seeds, with s.d.\ reported for F1. L+FH denotes listener + floor-holder conditioning; the final configuration isolates the floor-holder's audio in channel~2.} \label{tab:event_f1} \begin{tabular}{llrrlrrl} \toprule & & \multicolumn{3}{c}{Onset (horizon not compensated)} & \multicolumn{3}{c}{Window (compensated)} \\ \cmidrule(lr){3-5}\cmidrule(lr){6-8} Configuration & Slice & P & R & F1 & P & R & F1 \\ \midrule None & seen & 0.045 & 0.110 & $0.064 \pm 0.005$ & 0.073 & 0.179 & $0.103 \pm 0.005$ \\ & unseen & 0.029 & 0.082 & $0.042 \pm 0.003$ & 0.046 & 0.133 & $0.068 \pm 0.004$ \\ Listener & seen & 0.057 & 0.103 & $0.070 \pm 0.008$ & 0.091 & 0.163 & $0.112 \pm 0.012$ \\ & unseen & 0.029 & 0.083 & $0.041 \pm 0.005$ & 0.047 & 0.131 & $0.066 \pm 0.002$ \\ L+FH (mixed) & seen & 0.053 & 0.160 & $0.079 \pm 0.004$ & 0.076 & 0.227 & $0.112 \pm 0.004$ \\ & unseen & 0.030 & 0.120 & $0.047 \pm 0.003$ & 0.045 & 0.180 & $0.070 \pm 0.002$ \\ L+FH (FH ch2) & seen & 0.062 & 0.120 & $0.079 \pm 0.004$ & 0.095 & 0.181 & $0.120 \pm 0.005$ \\ & unseen & 0.032 & 0.086 & $0.045 \pm 0.001$ & 0.050 & 0.134 & $0.070 \pm 0.004$ \\ \bottomrule \end{tabular} \end{table} \section{Hyperparameters and compute} \label{app:hparams} Table~\ref{tab:hparams} summarizes the settings shared across all 19 runs: the grid of four configurations with three seeds each, together with the seven-run remedy matrix using seed 13. Wall-clock time is computed as the sum of the per-epoch times recorded for each run. Since jobs were run sequentially on a single device, wall-clock hours and GPU-hours are equivalent here. Evaluation and inference passes were not timed separately and are therefore not included. \begin{table}[htbp] \centering \small \caption{Training hyperparameters and compute.} \label{tab:hparams} \begin{tabular}{ll} \toprule Setting & Value \\ \midrule Optimizer & AdamW \\ Learning rate & $1\times10^{-4}$ \\ Weight decay & $0.01$ \\ Gradient clipping & norm $1.0$ (adversary clipped as its own group) \\ Batch size & 16 \\ Max epochs & 12 \\ Early stopping & patience 6 on validation BC AUPRC, min-delta $0.002$ \\ Seeds & 13, 21, 42 (remedy matrix: 13 only) \\ Frozen parameters & 4{,}345{,}884 ($2\times$ CPC encoder) \\ Trainable parameters & 3{,}611{,}908 (trunk) / 3{,}958{,}724 (with conditioning) \\ Adapter trainable parameters & 356K (adapters and heads) \\ \midrule GPU & NVIDIA Thor, $\times1$, jobs sequential \\ Mean wall-clock per grid run & 1.87~h \\ Grid total (12 runs) & 22.4~GPU-h \\ Remedy matrix total (7 runs) & 10.6~GPU-h \\ Total (19 runs) & 33.0~GPU-h \\ \bottomrule \end{tabular} \end{table}

\section{Broader impacts}
\label{sec:impact}
Backchannel prediction can help conversational agents and social robots act as more
natural listeners, making spoken interaction smoother and potentially more accessible.
Adapting to an individual's listening behavior, however, raises privacy and consent
questions: personalization should be transparent to the user and should not rely on
behavioral data collected without their knowledge. Our benchmark derives from the AMI
corpus, collected for research with participant consent. Our finding that backchannel
behavior varies a lot across individuals also cautions against assuming a
single model generalizes across users or populations; we release our splits and
evaluation tools so such differences can be measured directly.

\end{document}